\documentclass{article}
\pdfoutput=1
\usepackage{graphicx}
\usepackage{authblk} 
\usepackage{amsmath}
\usepackage{xcolor}
\usepackage{multicol}
\usepackage{geometry}
\usepackage{xr} 
\makeatletter
\newcommand*{\addFileDependency}[1]{
\typeout{(#1)}
\IfFileExists{#1}{}{\typeout{No file #1.}}
}\makeatother
\newcommand*{\myexternaldocument}[1]{%
\externaldocument{#1}%
\addFileDependency{#1.tex}%
\addFileDependency{#1.aux}%
}
\myexternaldocument{supplementary_material}

\usepackage[capitalise]{cleveref}

\usepackage{natbib}
\title{Do Slip-Weakening Laws Shapes Influence Rupture Dynamics?}

\author[1]{Roxane Ferry}
\author[1]{Jean-François Molinari}
\affil[1]{Institute of Civil Engineering, Institute of Materials Science and Engineering, École Polytechnique Fédérale de Lausanne (EPFL), Lausanne, Switzerland}

\date{}

\begin{document}

\maketitle

\begin{abstract} 
    \par To model rupture dynamics, a friction law must be assumed. Commonly used constitutive laws for modeling friction include slip-weakening laws which are characterized by a drop from static to dynamic frictional stress. Within this framework, the prevailing understanding asserts that the frictional behavior is solely controlled by the fracture energy -- the area beneath the frictional stress versus the cumulative slip curve. In particular, it is claimed that the curve's shape itself has no influence on the system's response. Here we perform fully dynamic rupture simulations to challenge prevailing beliefs by demonstrating that the constitutive law shape exerts an intimate control over rupture profiles. For a consistent fracture energy but varying constitutive law shapes, the velocity profile is different: each abrupt slope transition leads to the localization of a distinct velocity peak. For example, in the case of a bilinear slip-weakening law featuring two different slopes, the rupture exhibits two distinct velocity peaks. This phenomenon arises from the transition between a constant weakening rate to another. However, this distinction does not seem to influence how a rupture responds to a stress barrier or the cumulative radiated energy emitted. These results are derived through two separate numerical schemes (spectral boundary integral and finite element methods) ensuring their independence from the computational approach employed.
\end{abstract}

\twocolumn

\section{Introduction}

    \par Understanding the fundamental mechanisms governing rupture phenomena is crucial for predicting and mitigating associated hazards. Numerous studies have focused on elucidating the complexities of rupture dynamics across various scales, from stick-slip behavior at microscale interfaces to large-scale seismic events \citep{rubino_understanding_2017, ke_earthquake_2022, passelegue_initial_2020, roch_dynamic_2023, cattania_precursory_2021, cocco_fracture_2023}. For instance, investigations into fault mechanics have shed light on the propagation and arrest of seismic ruptures, revealing intricate interactions between frictional forces, stress distribution, fault geometry and external loading conditions \citep{barras_emergence_2019, lebihain_earthquake_2021, rezakhani_finite_2020, roch_velocity-driven_2022, lapusta_nucleation_2003, romanet_fast_2018, cebry_creep_2022}. A powerful framework to grasp frictional dynamics across different contexts is the analogy between frictional rupture and conventional fracture mechanics \citep{fineberg_recent_2015, svetlizky_brittle_2019, barras_emergence_2020, reches_earthquakes_2023, weng_integrated_2022, kammer_earthquake_2024}. This analogy serves as a valuable tool to describe the dynamics of frictional ruptures. Recent studies demonstrate its effectiveness in quantitatively explaining the dynamic propagation and arrest of ruptures, both in laboratory experiments and numerical simulations \citep{svetlizky_classical_2014, rubinstein_detachment_2004, kammer_linear_2015, bayart_fracture_2016, shi_how_2023, gvirtzman_nucleation_2021, ke_rupture_2018}. However, it is worth noting that some studies have pointed out the limitations of this analogy, particularly concerning rupture-related dissipation \citep{brener_unconventional_2021, paglialunga_frictional_2024} or nucleation at interfaces with heterogeneous properties \citep{castellano_nucleation_2023}.

    \par The role of friction in the dynamics of rupture is critical, yet understanding its precise nature remains a challenge. It necessitates the reliance on assumed friction laws to effectively model rupture phenomena. Among the constitutive laws aimed at describing interfacial response, cohesive laws are frequently employed. These laws define interfacial stress solely as a function of cumulative opening (slip), incorporating a drop from a static to a dynamic stress across a characteristic distance $D_c$ that is interpreted as the slip required to renew surface contacts. The manner in which this transition occurs is dictated by the particular law considered. It is widely held that the frictional behavior is determined exclusively by the fracture energy, which is to say, the area beneath the curve of frictional strength versus cumulative slip, rather than by the shape of the curve itself \citep{freund_dynamic_1998}.
    \par While the linear slip-weakening law is the most frequently used constitutive law, bilinear and multilinear laws (\cref{fig:geometry}b-d) have been used in earthquakes, composite material and soil simulations. \citet{paglialunga_scale_2022} employed a bilinear law to account for two different weakening mechanisms identified in stick-slip experiments. Similarly, following the law introduced by \citet{kanamori_microscopic_2000} to account for melting or pressurization during slip, \citet{galvez_rupture_2016} modeled the $M_w$ 9.0 Tohoku earthquake using a law with two sequential strength drops to account for rupture reactivation. \citet{barani_fracture_2016} have used bilinear laws to model cohesive soils with adhesive additives. In composite material simulations 
    \citep{petersson_crack_1981, li_use_2005, li_mixed-mode_2006, davila_procedure_2009, dourado_bilinear_2012, de_morais_bilinear_2015}, bilinear and multilinear laws have been used to account for fiber bridging and pull-out. \citet{li_mixed-mode_2006} underscored the adoption of a bilinear laws as an acknowledgment that the shape of the cohesive law can significantly influence the fracture behavior of interfaces. However, the impact of the cohesive law's shape, considered independently from fracture energy, has not been thoroughly investigated.
    
    \par In this study, we challenge prevailing beliefs by conducting fully dynamic rupture simulations, revealing the intimate control exerted by the constitutive law shape on rupture profiles. In the method section, we provide a comprehensive overview of the numerical framework utilized. Subsequently, we compare the influence of three distinct constitutive laws (linear, bilinear, and multilinear slip-weakening laws) on the velocity profile. Following this, we conduct a comparative analysis of the rupture dynamics in simulations employing either a linear or a bilinear law, while maintaining a consistent fracture energy, and we examine if these laws influence the response to a stress barrier. Finally, the discussion section analyzes the findings and their broader implications.

\section{Method}

    \par
\begin{figure}[h!]
        \centering
        \includegraphics[width=\linewidth]{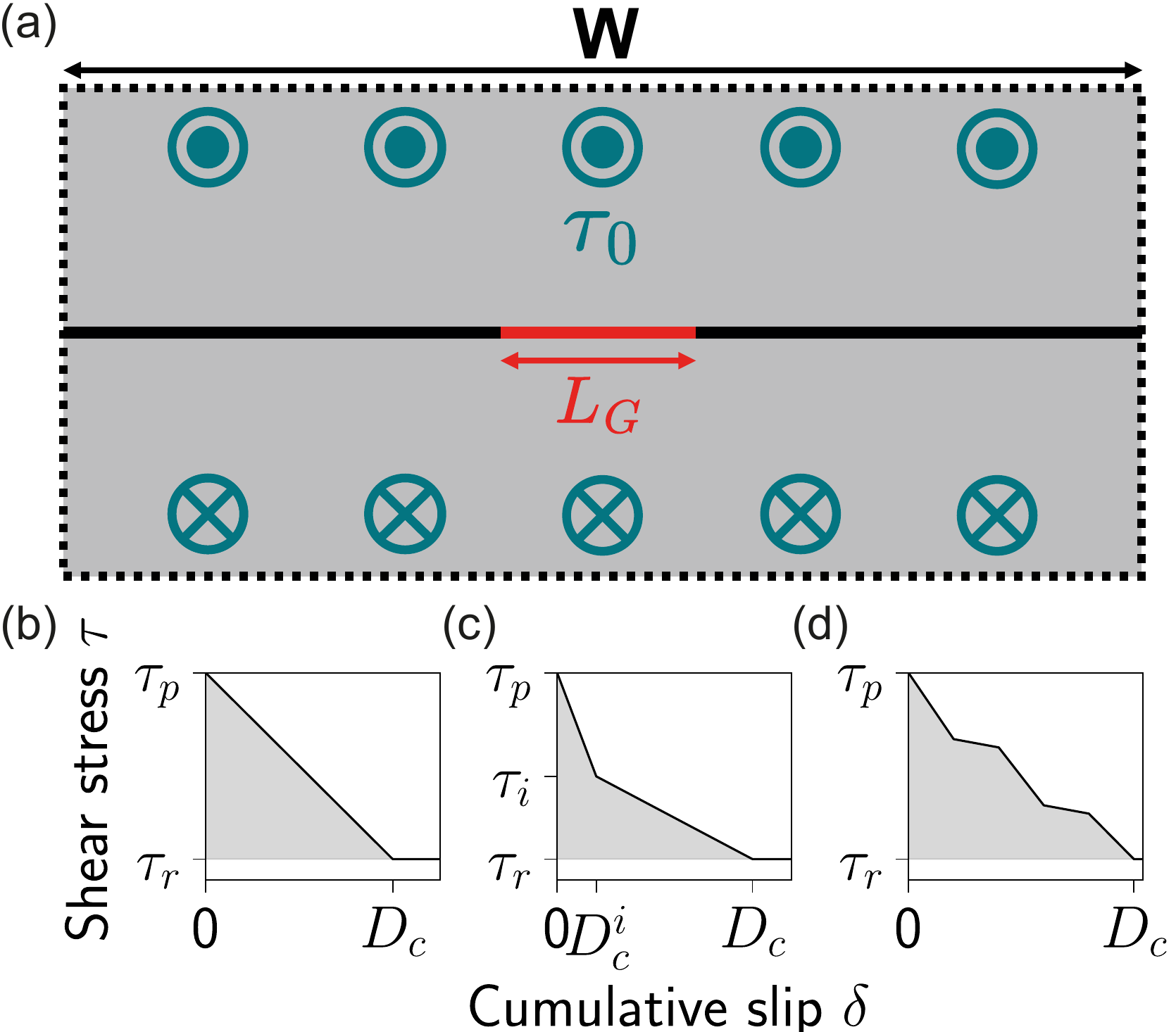}
        \caption{\textbf{(a)} Schematic representation of the system. Two elastic bodies (gray) are in contact along an interface (black solid line) of length $W$. The system is subjected to anti-plane shear (mode III) conditions through a far-field loading $\tau_0$. The red segment denotes the crack seed to promote nucleation. \textbf{(b, c, d)} Linear, bilinear and multilinear slip-weakening laws, respectively.}
        \label{fig:geometry}    
    \end{figure}

 We perform fully dynamic rupture simulations in 2D using the in-house open source spectral boundary integral formulation of the elastodynamic equations \textit{cRacklet} \citep{roch_cracklet_2022}. This method, originally proposed by \citet{geubelle_spectral_1995}, has been employed to accurately capture the intricate dynamics of rupture phenomena using FFT to mitigate computational costs. In this approach, only the interface is discretized. We consider two semi-infinite linearly elastic solids in contact along a planar interface of length $W$ loaded under anti-plane shear (mode III) conditions induced by a far-field stress $\tau_0$ (\cref{fig:geometry}a). 

    \par We employ various slip-weakening laws (\cref{fig:geometry}b-d) to describe the frictional behavior of the interface, aiming to compare their influence on the rupture profile. All these laws are characterized by a decrease from a peak shear stress $\tau_p$ to a residual shear stress $\tau_r$ as the cumulative slip $\delta$ increases over a characteristic distance $D_c$. This distance can be interpreted as the amount of slip necessary to renew surface contacts. We compare a linear slip-weakening law defined as \citep{palmer_growth_1973}:

    \begin{equation}
    	\tau = \left\{
    	\begin{array}{ll}
    		\tau_p - (\tau_r - \tau_p)\delta / D_c & \mbox{if } \delta \leq D_c \\
    		\tau_r & \mbox{if } \delta > D_c,
    	\end{array}
    	\right.
    	\label{eq:SW}
    \end{equation} 

    with a bilinear slip-weakening law:  

	\begin{equation}
		\tau = \left\{
		\begin{array}{ll}
			\tau_p - \left(\tau_p - \tau_i \right) \delta / D_c^i & \mbox{if } \delta \leq D_c^i, \\
			\tau_i - \left(\tau_i - \tau_r \right) \left(\delta - D_c^i\right) / \left(D_c - D_c^i\right)&  \mbox{if } D_c^i < \delta \leq D_c, \\
			\tau_r & \mbox{else}.
		\end{array}
		\right.
		\label{eq:BSW}
	\end{equation} 

    A multilinear slip-weakening law based on the same principle as the bilinear one but with five linear weakening stages is also used (\cref{fig:geometry}d). 
    
    \par Rupture is initiated by introducing a crack seed at the center of the interface, with the shear stress set to its residual value $\tau_r$. This crack seed has dimensions corresponding to Griffith's length, defined as $L_G = 4 G G_c / (\pi \tau_0^2)$, where $G$ is the shear modulus and $G_c$ the fracture energy. For all constitutive laws employed, the crack seed must exhibit dimensions of the Griffith's length for the rupture to initiate, in accordance with predictions of Linear Elastic Fracture Mechanics (LEFM) theory. Consequently, the different constitutive laws do not alter the nucleation process.
    \par To avoid any dynamic side effects at the start of the simulation, the far-field loading is gradually increased up to its final value $\tau_0$. 

    \par The findings presented in the following section have also been derived with the in-house open source finite element method \textit{Akantu} \citep{richart_akantu_2024}, ensuring their independence from the computational approach employed. Unlike the spectral boundary integral formulation, this approach involves the full discretization of solids in contact, allowing for interactions between waves and boundaries within the solids. Nevertheless, to facilitate comparison with the spectral boundary integral method, we ensured that the thickness was set sufficiently high to prevent waves from returning to the interface.

\section{Results}

\par All simulations herein are conducted using elastic bodies with material properties akin to PMMA (polymethyl methacrylate), featuring a Poisson's ratio $\nu$ of 0.33, a shear modulus $\mu$ of 3 GPa, and a density $\rho$ of 1200 kg.m\textsuperscript{-3}. The interface has a length of $30 \ L_G$ and is discretized using 20000 elements, ensuring an adequate resolution of the cohesive zone\footnote{The mechanical community prefers the term "process zone"} size $l_{cz} = 2 G G_c / \tau_p^2$ with more than 500 elements.

\subsection*{Influence of the constitutive law shape}

    \begin{figure*}[h!]
        \centering
        \includegraphics[width=\textwidth]{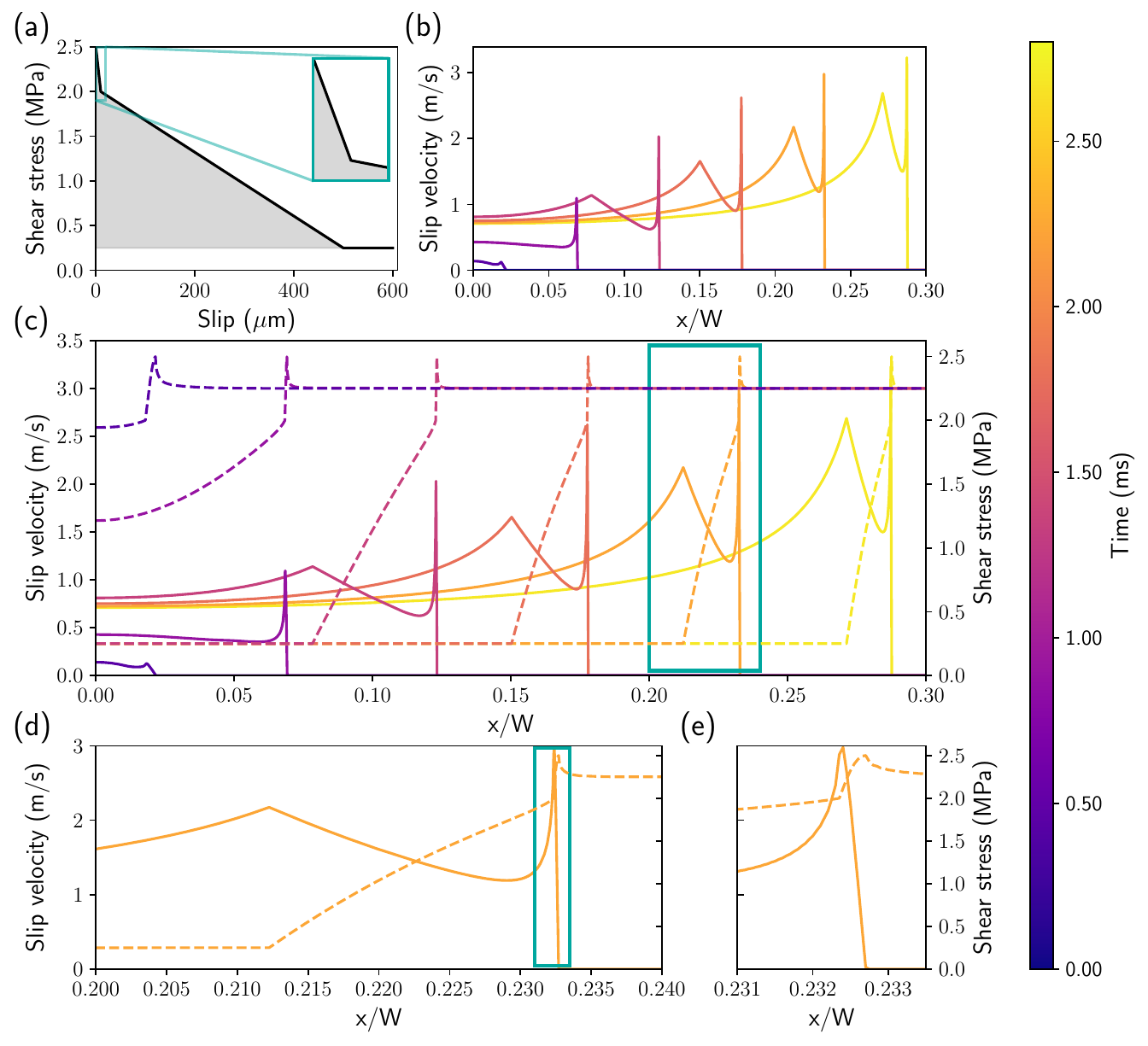}
        \caption{\textbf{(a)} Bilinear slip-weakening law used, with $\tau_p = 2.5$~MPa, $\tau_i = 2$~MPa, $\tau_r = 0.25$~MPa, $D_c = 5.10^{-4}$~m and $D_c^i = 10^{-5}$~m. The gray area represents the fracture energy $G_c = 448.75$ J.m\textsuperscript{2} \textbf{(b)} Slip velocity evolution along the right half of the interface of length W. The color represents time, from purple to yellow. \textbf{(c)} Same as (b) with the corresponding shear stress in dashed lines. The blue rectangle represents the zoomed-in region showcased in (d). \textbf{(d)} Zoom in on the two slip velocity peaks at a single time step, showcasing their alignment with the corresponding breaks in shear stress slopes. \textbf{(e)} Zoom in blue rectangle of (d).}
        \label{fig:bilinear}    
    \end{figure*}

    \begin{figure*}[h!]
        \centering
        \includegraphics[width=\textwidth]{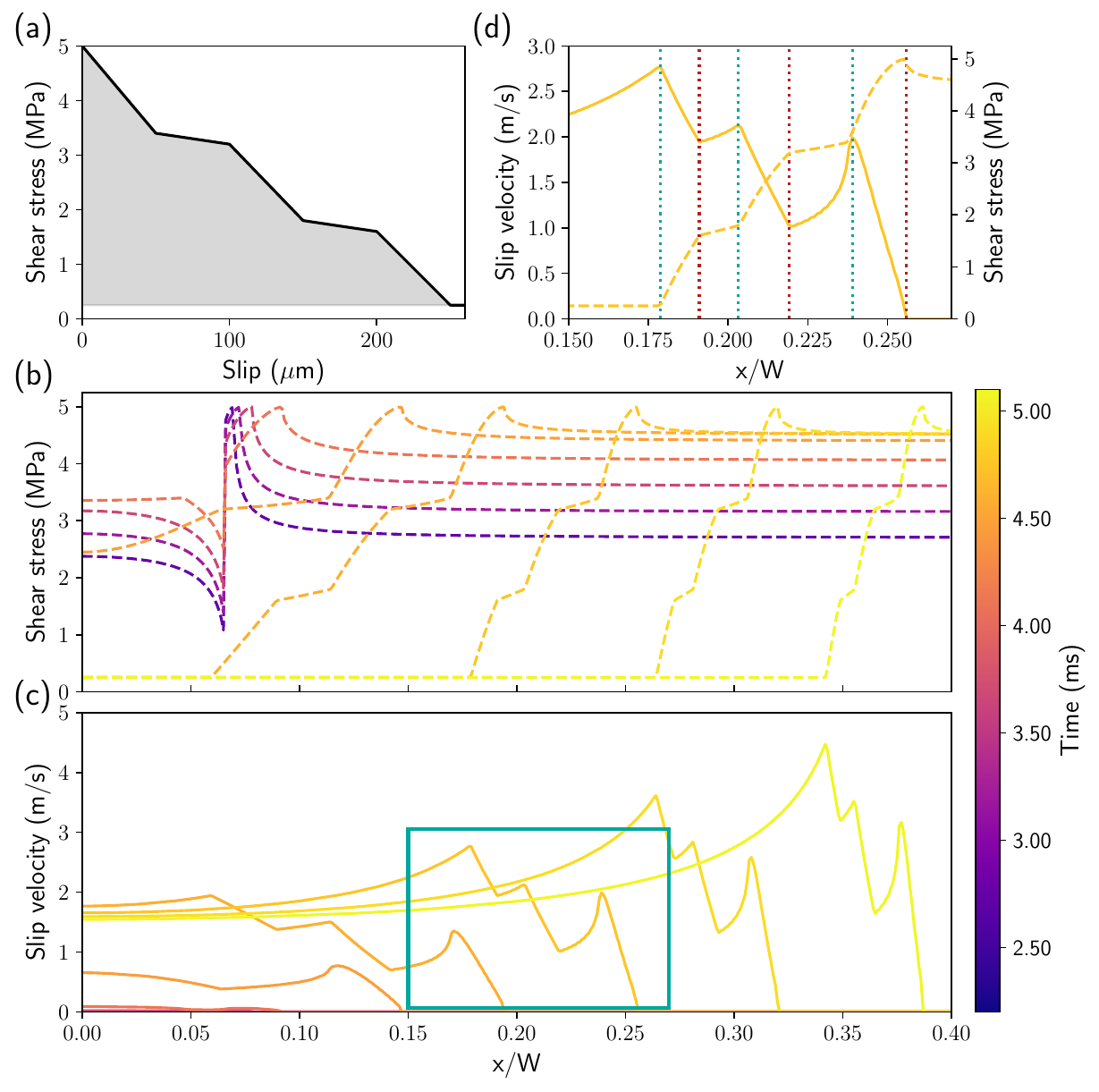}
        \caption{\textbf{(a)} Multilinear slip-weakening law used, with $\tau_p = 5$~MPa, $\tau_r = 0.25$~MPa, and the intermediate shear stresses are 3.4, 3.2, 1.8 and 1.6~MPa. $D_c = 25.10^{-5}$~m and the intermediate critical openings are $5.10^{-5}$, $10.10^{-5}$, $15.10^{-5}$ and $20.10^{-5}$~m respectively. \textbf{(b)} and \textbf{(c)} Shear stress evolution along the right half of the interface and the corresponding slip velocity evolution. The colorbar represents time, from purple to yellow. The blue rectangle represents the zoomed-in region showcased in (d). \textbf{(d)} Zoom in on a specific time step to highlight the three peaks in slip velocity, showcasing their alignment with the shear stress slope breaks coinciding with the sharpest slope gradients (blue dotted lines). Velocity troughs correspond with shear stress slope breaks, aligning with the more gradual slope gradients (vertical red dotted lines).}
        \label{fig:multilinear}    
    \end{figure*}

    \par Our goal is to investigate if the shape of the constitutive law influences the dynamics of rupture. In \cref{fig:bilinear} we illustrate the evolution of slip velocity profile during a dynamic rupture where the interface response is governed by a bilinear slip-weakening law (\cref{fig:bilinear}a). As the rupture propagates symmetrically from the center of the interface, only one half of the interface is depicted. A second peak emerges behind the rupture front (\cref{fig:bilinear}b). \cref{fig:bilinear}c showcases the evolution of slip velocity alongside corresponding shear stress. Remarkably, the velocity peaks precisely align with the shear stress slope discontinuities at all times (movies in SM). This behavior is further elucidated in \cref{fig:bilinear}d-e, where zoomed views of \cref{fig:bilinear}c are provided for a specific time step. During the initial stages, stress concentration at the crack tip intensifies with the amplification of the far-field loading, eventually leading to rupture initiation. Once the shear stress surpasses the peak shear stress $\tau_p$ (2.5 MPa in this case), slip initiation occurs. The first slip velocity peak corresponds to the location of the first shear stress slope discontinuity (located at $D_c^i$). In the subsequent phase of gradual weakening between $D_c^i$ and $D_c$, the slip velocity behind the primary peak increases, leading to the formation of a secondary slip velocity peak coinciding with the second slope discontinuity (at $D_c$). The emergence of this secondary peak aligns with previous numerical observations reported by \citet{paglialunga_scale_2022} employing the same constitutive law. Their findings also indicate that the secondary peak progressively approaches the primary one, eventually collapsing into a single peak, a phenomenon similarly observed in the later stages of our simulations (not shown here). Additionally, experimental evidence supporting the existence of such secondary peaks has been documented by \cite{berman_dynamics_2020}.
    
    \par This behavior is surprising because one might anticipate that the slip velocity profile would depend solely on the fracture energy, rather than the shape of the curve itself \citep{freund_dynamic_1998}. This phenomenon is also observed with linear slip weakening laws, where the singular velocity peak matches the shear stress slope discontinuity, though not presented here.

    \begin{figure*}[h!]
        \centering
        \includegraphics[width=\textwidth]{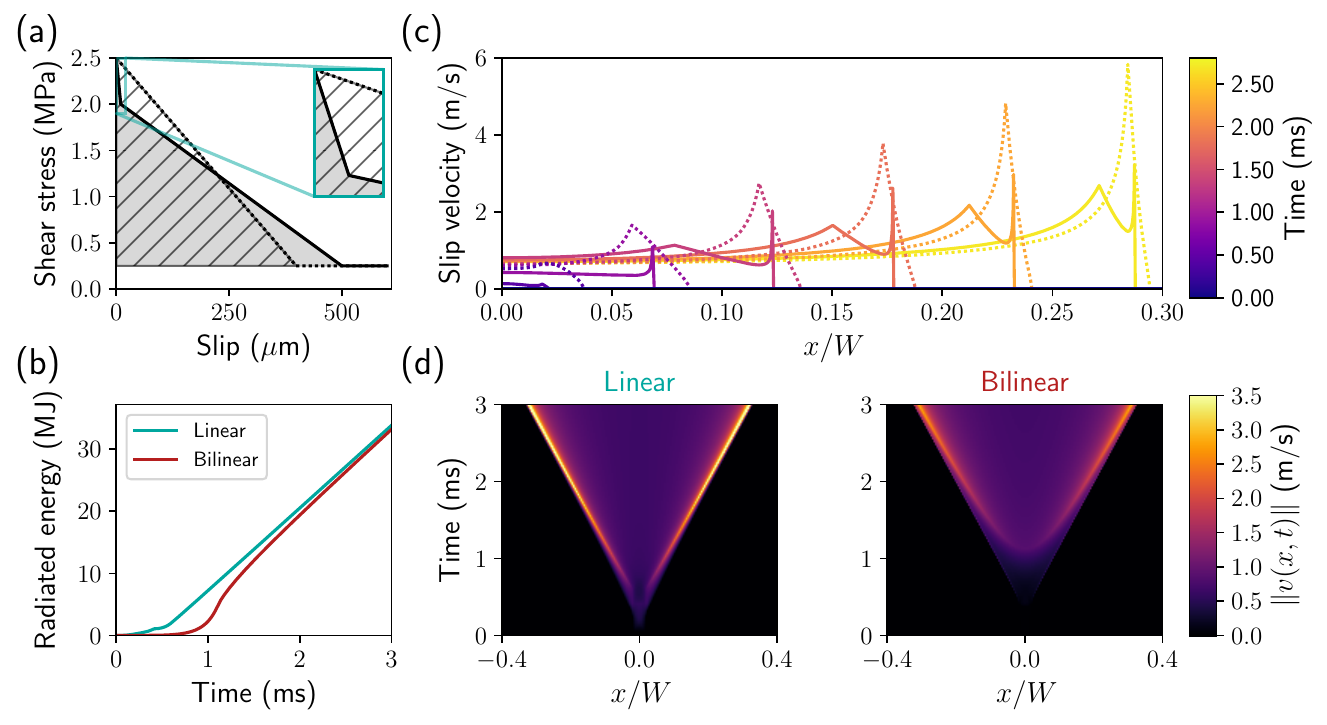}
        \caption{\textbf{(a)} Linear (dotted line) and bilinear (solid line) slip-weakening laws employed. Both laws have the same fracture energy depicted by the hatched and gray areas, respectively. \textbf{(b)} Comparison of radiated energy evolution during a rupture propagation with the two laws. \textbf{(c)} Comparison of slip velocity evolution along the right half of the interface of length W for the linear (dotted line) and bilinear (solid line) laws. The colorbar represents time, from purple to yellow. \textbf{(d)} Comparison of CT diagrams for the linear (left) and bilinear (right) laws. The colorbar represents the slip velocity from purple to yellow. Both reach quasi identical rupture speed.}
        \label{fig:comparison}    
    \end{figure*}

    \par The slip velocity profile is not limited to two peaks. We posit that, given the appropriate constitutive law, the slip profile can manifest any number of peaks. \cref{fig:multilinear} elucidates this concept through the depiction of slip velocity profile evolution governed by a multilinear slip-weakening law (\cref{fig:multilinear}a), which is characterized by five linear weakening stages. Notably, \cref{fig:multilinear}c depicts the presence of three distinct slip velocity peaks. Akin to the bilinear scenario, these velocity peaks align precisely with shear stress slope breaks (\cref{fig:multilinear}b) throughout the rupture process. To emphasize this correlation, \cref{fig:multilinear}d provides a detailed examination at a specific time step, revealing the correspondence between the slip velocity and shear stress profiles. The vertical blue dotted lines map the velocity peaks to the sharpest shear stress slope breaks, while the vertical red dotted lines denote the correlation between slip velocity troughs and shear stress slope discontinuities with more gradual gradients. This intricate correlation underscores the assertion that the slip velocity profile is entirely dictated by the chosen constitutive law, determining not only the number, but also the precise location and width of these peaks.

    \par This phenomenon can be elucidated as follows: each linear weakening stage corresponds to a specific constant rate of weakening. Consequently, an abrupt slope transition signifies a sudden change in weakening rate. In Figure \ref{fig:multilinear}d, the blue dotted lines represent transitions from a higher weakening rate to a lower one. At these transition points, the slip velocity decelerates, resulting in the peak being positioned at the transition location. Conversely, the red dotted lines denote transitions from a lower to a higher weakening rate, leading to an acceleration in slip velocity and the consequent appearance of velocity troughs at the transition locations.

    \par We have shown that slip velocity peaks occur along breaks in the constitutive law's slope. The bilinear and multilinear laws employed are $C^0$ continuous functions. How might the slip velocity profile appear with a constitutive law that maintains at least $C^1$ continuity, such as the hyperbolic tangent? In such a scenario, due to time discretization, regardless of how small the time step is, the constitutive law is inevitably sampled, leading to sharp transitions and consequently the emergence of peaks in slip velocity (see Fig.~S1).

\subsection*{Comparison at constant fracture energy}

    \begin{figure}[h!]
        \centering
        \includegraphics[width=\linewidth]{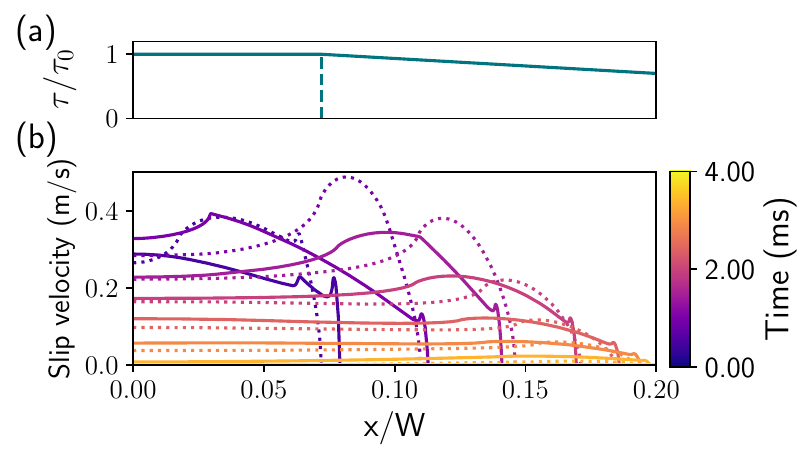}
        \caption{Comparison of the response to a stress barrier with the linear and bilinear slip-weakening laws. \textbf{(a)} Far-field loading profile along the interface showing a constant load until approximately x = 0.09 W (marked by the dashed line), followed by a gradual decrease towards 0. \textbf{(b)} Comparison of slip velocity evolution along the right half of the interface of length W for the linear (dotted line) and bilinear (solid line) laws. The colorbar represents time, from purple to yellow. The rupture governed by a bilinear slip-weakening law extends slightly beyond that governed by the linear law, by approximately 0.75 times the cohesive zone size, which is considered negligible.}
        \label{fig:barrier}    
    \end{figure}

    \par Our simulations reveal that the constitutive law's shape governs the slip velocity profile, challenging the prevailing notion that fracture energy alone dictates a system's rupture behavior. To disentangle the effects of fracture energy and curve shape, we compare two cases with identical fracture energy: one employing a linear slip-weakening law and the other the bilinear law previously used (\cref{fig:comparison}a). \cref{fig:comparison}c compares the evolution of slip velocity for both laws. As expected, the linear law results in a slip velocity profile characterized by a singular, pronounced peak, which exceeds the heights of the peaks observed under the bilinear law. In the linear law scenario, rupture initiates slightly earlier than in the bilinear law scenario, attributed to a higher average weakening rate in comparison to the bilinear law case. Despite these differences, both scenarios result in quasi identical rupture speeds (\cref{fig:comparison}d). The cumulative radiated energy is depicted in \cref{fig:comparison}b. In the early stages, the radiated energy in the linear case surpasses that of the bilinear case, owing to the higher slip velocity and rupture speed. However, as both ruptures reach a consistent rupture speed in later stages, the cumulative radiated energy ultimately becomes equivalent. Given that the fracture energy is the same for both cases, this conclusively demonstrates that it is indeed the shape of the constitutive law that controls the dynamics of the rupture process. 

    \medskip
    
    \par The outcomes of two rupture simulations, each characterized by identical fracture energy yet differing constitutive law shapes, yield distinctive velocity profiles. How does this disparity impact the response to a stress barrier? To evaluate this, we conduct rupture simulations in the presence of a stress barrier. The far-field loading is maintained constant around the central part of the interface and subsequently diminishes linearly towards zero (\cref{fig:barrier}a). This setup initiates the rupture under a uniform driving force, which is then progressively decreased. It contrasts with the work of \citet{paglialunga_scale_2022}, where they investigated the response to a stress barrier using linear and bilinear laws characterized by differing fracture energies to account for different weakening mechanisms.    
    \par The slip velocity profile evolution for the two laws is compared in \cref{fig:barrier}b. Both ruptures exhibit rapid decay as the driving force is decreased, with the rupture governed by the bilinear law extending marginally further than that governed by the linear law, by approximately 0.75 times the cohesive zone size $l_{cz}$. Given that the cohesive zone size represents the lengthscale at which dynamic processes occur, this additional distance is considered negligible. It appears that variations in slip velocity profiles, under constant fracture energy, do not significantly influence the response to a stress barrier. It is important to emphasize that the observed lack of impact of different slip-weakening laws on response to a stress barrier is attributed to their identical fracture energies. 

\section{Discussion and conclusion}

    \par We conducted fully dynamic rupture simulations to elucidate the influence of the constitutive law's shape on the characteristics of the rupture profiles, specifically its number of peaks, their spatial distribution, and their widths. Our investigation encompassed linear, bilinear, and multilinear slip-weakening laws, demonstrating that each abrupt slope transition in the constitutive law corresponds to the emergence of a localized peak in the slip velocity profile. For instance, in the case of a bilinear law, employed in studies of earthquakes or composite materials, a second peak manifests behind the primary one. This occurs due to a transition from one constant rate of weakening to another.
    \par Furthermore, our comparative analysis between linear and bilinear laws, while maintaining a constant fracture energy, underscores the pivotal role of constitutive law shapes in dictating rupture profiles independently of fracture energy considerations. Nevertheless, it is essential to highlight that fracture energy remains a critical factor governing nucleation behavior and the response to stress barriers. Our investigation thus provides clear evidence that while the fracture energy is crucial in dictating certain aspects of rupture dynamics, the specific features of the rupture profile are fundamentally shaped by the geometry of the constitutive law. This finding challenges the prevailing paradigm, which posits that the rupture profile is controlled by fracture energy alone \citep{freund_dynamic_1998}. This conventional understanding stems from analyses on steady-state crack growth scenarios. The discrepancy in behavior observed in our case arises from the fact that we are dealing with transient, rather than steady-state, crack growths.

    \par We emphasize that, considering the significant influence exerted by the arbitrary selection of the slip-weakening law, caution is warranted in the utilization of such constitutive laws. The dependency of the slip velocity profile on the constitutive law shape has significant implications for earthquake simulations, particularly when a slip-weakening restrengthening law is employed to generate pulse-like ruptures. This law involves a linear decrease of shear stress towards a residual value, followed by a linear increase, resulting in two abrupt slope transitions that define the pulse size. 
    Additionally, the presence of multiple slip velocity peaks introduces further challenges in identifying the crack-tip of a rupture, as well as in fitting a K-field.

\bibliography{SWPaper}

\end{document}


\maketitle

\section*{Hyperbolic tangent cohesive law}

\begin{figure}[h!]
    \centering
    \includegraphics[width=8cm]{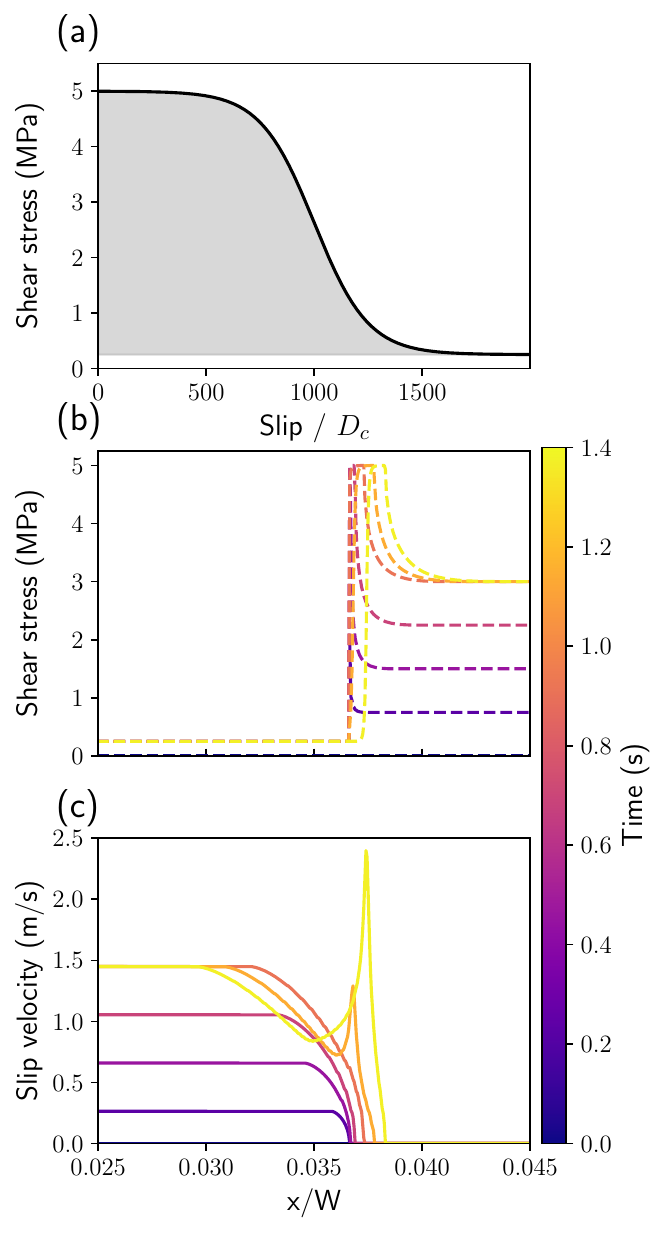}
    \caption{\textbf{(a)} Hyperbolic tangent slip-weakening law. \textbf{(b)} and \textbf{(c)} Shear stress evolution along the right half of the interface and the corresponding slip velocity evolution. The colorbar represents time, from purple to yellow.}
    \label{fig:SM_tanh}
\end{figure}

\par We used a hyperbolic tangent function to define a smooth cohesive law without any discontinuities in its derivative. The shear stress is given by:
\begin{equation}
    \tau = \frac{1}{2} \left(\tau_p + \tau_r\right) + \frac{1}{2} \left(\tau_p - \tau_r\right) \tanh \left(\frac{\text{center} - \delta / D_c}{\text{smoothing}} \right),
\end{equation}
where $\tau_p$ and $\tau_r$ are the peak and residual shear stresses, respectively, $\delta$ is the cumulative slip, $D_c$ is a characteristic distance and center and smoothing are parameters that adjust the shape of the hyperbolic tangent function.

\par As can been seen in \cref{fig:SM_tanh}, due to the time discretization, the law is inevitably sampled, which eventually leads to the emergence of a peak in the slip velocity. Therefore, the emergence of a slip velocity peak cannot be avoided.

\section*{Dugdale cohesive law}

\begin{figure}[h!]
    \centering
    \includegraphics[width=15cm]{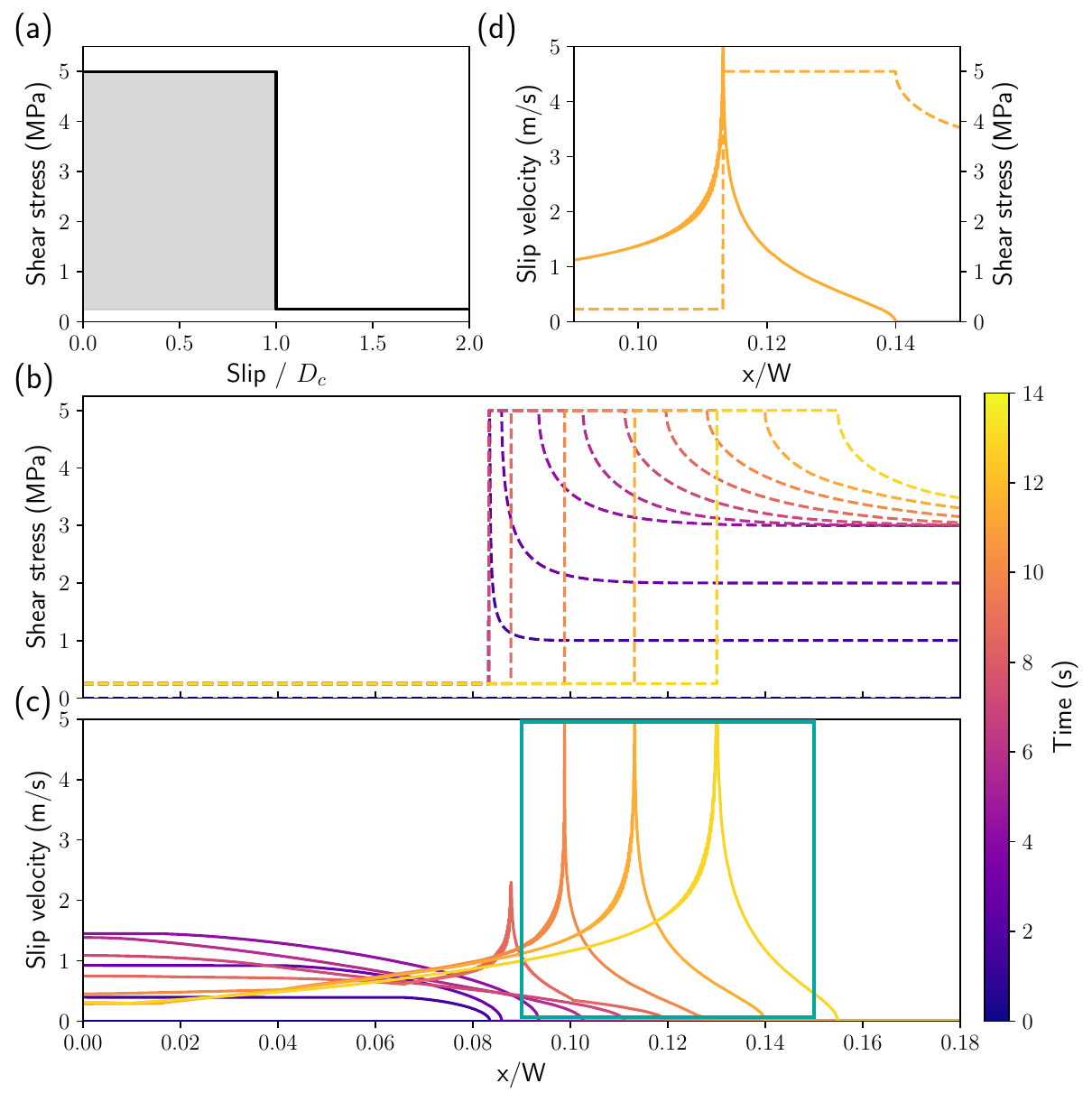}
    \caption{\textbf{(a)} Dugdale slip-weakening law. \textbf{(b)} and \textbf{(c)} Shear stress evolution along the right half of the interface and the corresponding slip velocity evolution. The colorbar represents time, from purple to yellow. The blue rectangle represents the zoomed-in region showcased in (d). \textbf{(d)} Zoom in on a specific time step to highlight the alignment of the slip velocity peak and the shear stress slope break.}
    \label{fig:SM_dugdale}
\end{figure}

\par In the case of a Dugdale law defined as:

\begin{equation}
    \tau = \left\{
    \begin{array}{ll}
        \tau_p & \mbox{if } \delta \leq D_c \\
        \tau_r & \mbox{if } \delta > D_c,
    \end{array}
    \right.
    \label{eq:SM_dugdale}
\end{equation} 
a slip velocity peak also emerges and localizes at $\delta = D_c$ (\cref{fig:SM_dugdale}).